\documentclass[10pt,a4paper,british]{article}
\usepackage[T1]{fontenc}
\usepackage[latin1]{inputenc}
\usepackage{amsmath}
\usepackage{amssymb}
\usepackage{setspace}
\usepackage{esint}
\makeatletter

\allowdisplaybreaks

\makeatother

\usepackage{babel}

\date{}

\makeatother

\usepackage{babel}
\begin{document}

\title{Spectrum generating algebra for the pure spinor superstring}

\author{Renann Lipinski Jusinskas%
\thanks{renannlj@ift.unesp.br%
}}

\maketitle

\begin{center}
ICTP South American Institute for Fundamental Research \\
Instituto de F\'isica Te\'orica, UNESP - Univ. Estadual Paulista \\
Rua Dr. Bento T. Ferraz 271, 01140-070, S\~ao Paulo, SP, Brasil.
\par\end{center}

\

\begin{abstract}
In this work, a supersymmetric DDF-like construction within the pure
spinor formalism is presented. Starting with the light-cone massless
vertices, the creation/annihilation algebra is derived in a simple
manner, enabling a systematic construction of the physical vertex
operators at any mass level in terms of $SO\left(8\right)$ superfields,
in both integrated and unintegrated forms.
\end{abstract}

\section*{Introduction}

The pure spinor formalism debuted more than a decade ago \cite{Berkovits:2000fe}
and remains the only one that allows Lorentz covariant computations
with manifest supersymmetry. Its fundamental piece is a BRST-like
charge
\begin{equation}
Q_{\textrm{BRST}}=\frac{1}{2\pi i}\oint\left(\lambda^{\alpha}d_{\alpha}\right)
\end{equation}
 that is nilpotent whenever $\lambda^{\alpha}$ is a pure spinor ($\lambda\gamma^{m}\lambda=0$).
Here,
\begin{eqnarray}
d_{\alpha} & = & p_{\alpha}+\frac{1}{2}\partial X^{m}\left(\gamma_{m}\theta\right)_{\alpha}+\frac{i}{8}\left(\theta\gamma^{m}\partial\theta\right)\left(\gamma_{m}\theta\right)_{\alpha},\\
\Pi^{m} & = & \partial X^{m}+\frac{i}{2}\left(\theta\gamma^{m}\partial\theta\right),
\end{eqnarray}
are the usual invariants defined by the supersymmetry charge
\begin{equation}
q_{\alpha}=\frac{1}{2\pi}\oint\left\{ -p_{\alpha}+\frac{1}{2}\partial X^{m}\left(\gamma_{m}\theta\right)_{\alpha}+\frac{i}{24}\left(\theta\gamma^{m}\partial\theta\right)\left(\gamma_{m}\theta\right)_{\alpha}\right\} ,
\end{equation}
and satisfy\begin{subequations}
\begin{eqnarray}
\Pi^{m}\left(z\right)\Pi^{n}\left(y\right) & \sim & -\frac{\eta^{mn}}{\left(z-y\right)^{2}},\\
d_{\alpha}\left(z\right)\Pi^{m}\left(y\right) & \sim & -\frac{\gamma_{\alpha\beta}^{m}\partial\theta^{\beta}}{\left(z-y\right)},\\
d_{\alpha}\left(z\right)d_{\beta}\left(y\right) & \sim & i\frac{\gamma_{\alpha\beta}^{m}\Pi_{m}}{\left(z-y\right)}.
\end{eqnarray}
\end{subequations}As usual, $m=0,\ldots,9$ and $\alpha=1,\ldots,16$
are the spacetime vector and chiral spinor indices, respectively.
Observe that the fundamental length of the string is being fixed through
$\alpha'=2$ but it can be easily recovered by dimensional analysis.

In spite of its unknown origin, the power of the formalism resides
on the simple form of $Q_{\textrm{BRST}}$, which enables an elegant
treatment of the cohomology in terms of superfields. For example,
the massless open superstring states are in the cohomology of $Q_{\textrm{BRST}}$
at ghost number one, represented by $U=\lambda^{\alpha}A_{\alpha}\left(X,\theta\right)$.
Denoting the supersymmetric derivative as $D_{\alpha}=i\partial_{\alpha}-\frac{1}{2}\left(\gamma_{\alpha\beta}^{m}\theta^{\beta}\right)\partial_{m}$,
the condition $\left\{ Q_{\textrm{BRST}},U\right\} =0$ implies $D\gamma^{mnpqr}A=0$,
which is the linearized equation of motion for the super Yang-Mills
field $A_{\alpha}$. The gauge transformation $\delta A_{\alpha}=D_{\alpha}\Lambda$
is trivially reproduced in terms of BRST-exact states. The integrated
form of the vertex $U$ is given by

\begin{equation}
V=\frac{1}{2\pi i}\oint\left\{ \Pi^{m}A_{m}+i\partial\theta^{\alpha}A_{\alpha}+id_{\alpha}W^{\alpha}+N^{mn}F_{mn}\right\} ,\label{eq:integratedmassless}
\end{equation}
where the superfields $A_{m}$, $W^{\alpha}$ and $F_{mn}$ are defined
as\begin{subequations}\label{eq:superfields}
\begin{eqnarray}
A_{m} & \equiv & \frac{1}{8i}\left(D_{\alpha}\gamma_{m}^{\alpha\beta}A_{\beta}\right),\label{eq:supervector}\\
\left(\gamma_{m}W\right)_{\alpha} & \equiv & \left(D_{\alpha}A_{m}+\partial_{m}A_{\alpha}\right),\label{eq:superspinor}\\
F_{mn} & \equiv & \frac{1}{2}\left(\partial_{m}A_{n}-\partial_{n}A_{m}\right)\\
 & = & \frac{i}{16}\left(\gamma_{mn}\right)_{\phantom{\alpha}\beta}^{\alpha}D_{\alpha}W^{\beta}.\nonumber
\end{eqnarray}
\end{subequations}

Both $U$ and $V$ were successfully used in loop amplitude computations
and the results were shown to agree with the RNS superstring up to
two loops \cite{Berkovits:2005ng}. On the other hand, the construction
of the massive states is quite hard and only the first level of the
open superstring has been studied in detail \cite{Berkovits:2002qx}.
In fact, the proof that the pure spinor cohomology is equivalent to
the light-cone Green-Schwarz spectrum was obtained in \cite{Berkovits:2000nn}
through a complicated procedure, where the pure spinor variable was
written in terms of $SO\left(8\right)$ variables, involving an infinite
chain of ghost-for-ghosts. Later, the equivalence of the pure spinor
spectrum with the traditional superstring formalisms was demonstrated
in different ways \cite{Berkovits:PSRNSGS}, involving field redefinitions
and similarity transformations, but an explicit superfield description
of the massive states was still lacking.

Inspired on the work of Del Giudice, Di Vecchia and Fubini (DDF) for
the bosonic string \cite{Del Giudice:1971fp}, this work presents
a generalization of the spectrum generating algebra to the pure spinor
superstring. A DDF construction within the pure spinor formalism was
already discussed in \cite{Mukhopadhyay:2005zu}. However, the approach
of Mukhopadhyay has two main differences. First, the gauge (Wess-Zumino)
used there to build the DDF operators introduces unusual terms in
the creation/annihilation algebra, demanding an extended argument
for proving the validity of the construction, related to the orthonormality
of the transverse Hilbert space. Second, the lack of an explicit expression
for the DDF operators in \cite{Mukhopadhyay:2005zu}, although sufficient
for studying the D-Brane boundary states, makes the superfield description
of the massive states incomplete. Here, a way to systematically obtain
the pure spinor vertex operators at any mass level in terms of $SO\left(8\right)$
superfields will be presented. Starting with the massless states in
a particular light-cone gauge, which renders the massless vertices
independent of half of the $\theta^{\alpha}$ components, the spectrum
generating algebra will be shown to reproduce the expected superstring
spectrum. As a by-product of this construction, Siegel's proposal
for the massless superstring vertex \cite{Siegel:1985xj} will be
shown to give origin to a tachyonic state, reinforcing its quantum
inequivalence with the massless vertex of the RNS string. In parallel,
some particularities of the pure spinor approach will be discussed.

\section*{$SO\left(8\right)$ superfields}

It will be useful to establish the notation used here for the $SO\left(8\right)$
decomposition. For any $SO\left(1,9\right)$ vector $N^{m}$, the
light-cone directions are represented by $\sqrt{2}N^{\pm}\equiv\left(N^{0}\pm N^{9}\right)$
while the remaining spatial directions will be written as $N^{i}$,
with $i=1,\ldots,8$. The scalar product between $N^{m}$ and $P^{m}$
is simply $N^{m}P_{m}=-N^{+}P^{-}-N^{-}P^{+}+N^{i}P_{i}$. For a rank-$2$
antisymmetric tensor $N^{mn}$, the $SO\left(8\right)$ components
are defined to be $N^{ij}$, $N^{i}\equiv N^{-i}$, $\overline{N}^{i}\equiv N^{+i}$
and $N\equiv N^{+-}$. The map between $SO\left(1,9\right)$ and $SO\left(8\right)$
spinor indices is given generically by
\begin{eqnarray*}
\xi^{\alpha}=P_{a}^{\alpha}\xi^{a}+P_{\dot{a}}^{\alpha}\xi^{\dot{a}}, & \xi^{a}=P_{\alpha}^{a}\xi^{\alpha}, & \xi^{\dot{a}}=P_{\alpha}^{\dot{a}}\xi^{\alpha},\\
\chi_{\alpha}=P_{\alpha}^{a}\chi_{a}+P_{\alpha}^{\dot{a}}\chi_{\dot{a}}, & \chi_{a}=P_{a}^{\alpha}\chi_{\alpha}, & \chi_{\dot{a}}=P_{\dot{a}}^{\alpha}\chi_{\alpha},
\end{eqnarray*}
where the $SO\left(8\right)$ spinor indices are $a,b,\ldots$ and
$\dot{a},\dot{b},\ldots$ (chiral and antichiral, respectively) running
from $1$ to $8$, and $\left\{ P_{a}^{\alpha},P_{\dot{a}}^{\alpha},P_{\alpha}^{a},P_{\alpha}^{\dot{a}}\right\} $
form a complete basis of projectors satisfying
\[
\begin{array}{cc}
P_{a}^{\alpha}P_{\alpha}^{b}=\delta_{a}^{b}, & P_{\dot{a}}^{\alpha}P_{\alpha}^{\dot{b}}=\delta_{\dot{a}}^{\dot{b}},\\
P_{a}^{\alpha}P_{\alpha}^{\dot{a}}=0, & \delta_{\beta}^{\alpha}=P_{a}^{\alpha}P_{\beta}^{a}+P_{\dot{a}}^{\alpha}P_{\beta}^{\dot{a}},
\end{array}
\]
In this language, the pure spinor constraint is translated to
\begin{equation}
\lambda^{a}\lambda_{a}=\lambda^{\dot{a}}\lambda_{\dot{a}}=\lambda^{a}\lambda^{\dot{a}}\sigma_{a\dot{a}}^{i}=0,\label{eq:PS-SO8}
\end{equation}
where $\sigma_{a\dot{a}}^{i}$ are the $SO\left(8\right)$ analogues
of the Pauli matrices and satisfy\begin{subequations}\label{eq:so8gammacommutations}
\begin{eqnarray}
\sigma_{a\dot{a}}^{i}\sigma_{b\dot{a}}^{j}+\sigma_{b\dot{a}}^{i}\sigma_{a\dot{a}}^{j} & = & 2\eta^{ij}\eta_{ab},\\
\sigma_{a\dot{a}}^{i}\sigma_{a\dot{b}}^{j}+\sigma_{b\dot{a}}^{i}\sigma_{a\dot{a}}^{j} & = & 2\eta^{ij}\eta_{\dot{a}\dot{b}},\\
\sigma_{a\dot{a}}^{i}\sigma_{b\dot{b}}^{i}+\sigma_{b\dot{a}}^{i}\sigma_{a\dot{b}}^{i} & = & 2\eta_{ab}\eta_{\dot{a}\dot{b}}.
\end{eqnarray}
\end{subequations}Here, $\eta_{ij}$, $\eta_{ab}$ and $\eta_{\dot{a}\dot{b}}$
are the $SO\left(8\right)$ metrics of the vector and spinor indices.
Since they are flat metrics with $+$ signature, upper and lower $SO\left(8\right)$
indices will not be distinguished in this work. Finally, the $SO\left(1,9\right)$
gamma matrices will be written as
\[
\begin{array}{rclrcl}
\left(\gamma^{i}\right)^{\alpha\beta} & \equiv & P_{a}^{\alpha}\sigma_{a\dot{a}}^{i}P_{\dot{a}}^{\beta}+P_{a}^{\beta}\sigma_{a\dot{a}}^{i}P_{\dot{a}}^{\alpha}, & \left(\gamma^{i}\right)_{\alpha\beta} & \equiv & P_{\alpha}^{a}\sigma_{a\dot{a}}^{i}P_{\beta}^{\dot{a}}+P_{\beta}^{a}\sigma_{a\dot{a}}^{i}P_{\alpha}^{\dot{a}},\\
\left(\gamma^{-}\right)^{\alpha\beta} & \equiv & \sqrt{2}P_{a}^{\alpha}P_{a}^{\beta}, & \left(\gamma^{-}\right)_{\alpha\beta} & \equiv & -\sqrt{2}P_{\alpha}^{\dot{a}}P_{\beta}^{\dot{a}},\\
\left(\gamma^{+}\right)^{\alpha\beta} & \equiv & \sqrt{2}P_{\dot{a}}^{\alpha}P_{\dot{a}}^{\beta}, & \left(\gamma^{+}\right)_{\alpha\beta} & \equiv & -\sqrt{2}P_{\alpha}^{a}P_{\beta}^{a},
\end{array}
\]
and can be shown to satisfy $\left\{ \gamma^{m},\gamma^{n}\right\} =2\eta^{mn}$.
Observe that the spinor projectors are being defined implicitily so
their explicit form will never be required.

In the construction to be presented, left and right-moving fields
will be split. The only subtlety comes from the worldsheet scalars
$X^{m}$, which will be written as
\begin{equation}
X^{m}\left(z,\overline{z}\right)=X_{L}^{m}\left(z\right)+X_{R}^{m}\left(\overline{z}\right).
\end{equation}
This will be useful in studying holomorphic and anti-holomorphic sectors
in an independent manner. The difference between open and closed strings
will be discussed later, in the analysis of the spectrum.

Suppose one starts analyzing the unintegrated massless vertex with
a definite momentum $P^{m}=\frac{1}{2\pi}\oint\partial X^{m}$. From
the Lorentz covariance of the theory, an equivalent state can be constructed
with momentum $P^{+}=\sqrt{2}k$ and $P^{-}=0$ in another frame.
The gauge transformation can be used to set $A_{\dot{a}}$ to zero
and to eliminate the $\theta_{a}$-dependence \cite{Berkovits:2014bra},
in such a way that $\lambda^{\alpha}A_{\alpha}$ can be rewritten
as $\lambda^{a}A_{a}=a^{i}U_{i}+\chi^{a}Y_{a}$, where $a^{i}$ and
$\chi^{a}$ are the polarizations of the massless vector and spinor,
respectively, and\begin{subequations}\label{eq:masslessverticesp+}
\begin{eqnarray}
U_{i}\left(z;k\right) & \equiv & e^{-ik\sqrt{2}X_{L}^{-}}\left\{ \overline{\Lambda}_{i}+\left(\frac{k}{3!}\right)\overline{\theta}_{ij}\overline{\Lambda}_{j}+\left(\frac{k^{2}}{5!}\right)\overline{\theta}_{ij}\overline{\theta}_{jk}\overline{\Lambda}_{k}+\left(\frac{k^{3}}{7!}\right)\overline{\theta}_{ij}\overline{\theta}_{jk}\overline{\theta}_{kl}\overline{\Lambda}_{l}\right\} ,\\
Y_{a}\left(z;k\right) & \equiv & e^{-ik\sqrt{2}X_{L}^{-}}\left(\sigma^{i}\theta\right)_{a}\left\{ \left(\frac{1}{2!}\right)\overline{\Lambda}_{i}+\left(\frac{k}{4!}\right)\overline{\theta}_{ij}\overline{\Lambda}_{j}+\left(\frac{k^{2}}{6!}\right)\overline{\theta}_{ij}\overline{\theta}_{jk}\overline{\Lambda}_{k}+\left(\frac{k^{3}}{8!}\right)\overline{\theta}_{ij}\overline{\theta}_{jk}\overline{\theta}_{kl}\overline{\Lambda}_{l}\right\} \nonumber \\
 &  & +e^{-ik\sqrt{2}X_{L}^{-}}\left(\frac{\lambda_{a}}{k}\right).
\end{eqnarray}
\end{subequations}Here, $\overline{\Lambda}^{i}\equiv\left(\lambda^{a}\sigma_{a\dot{a}}^{i}\theta^{\dot{a}}\right)$,
$\overline{\theta}^{ij}\equiv\left(\theta^{\dot{a}}\sigma_{\dot{a}\dot{b}}^{ij}\theta^{\dot{b}}\right)$
and $\sigma^{ij}\equiv\sigma^{[i}\sigma^{j]}$. To show the BRST-closedness
of \eqref{eq:masslessverticesp+} note that
\begin{eqnarray}
\left[Q_{\textrm{BRST}},\overline{\theta}^{ij}\right] & = & 2i\overline{\Lambda}^{ij},\\
\left(\sigma_{i}\lambda\right)_{a}\overline{\theta}^{ij} & = & -\left(\sigma_{i}\theta\right)_{a}\overline{\Lambda}^{ij}+3\overline{\Lambda}\left(\sigma^{j}\theta\right)_{a},\\
\overline{\Lambda}_{ij}\overline{\Lambda}_{j} & = & -3\overline{\Lambda}_{i}\overline{\Lambda},
\end{eqnarray}
with $\overline{\Lambda}\equiv\left(\lambda^{\dot{a}}\theta_{\dot{a}}\right)$
and $\overline{\Lambda}^{ij}\equiv\left(\lambda^{\dot{a}}\sigma_{\dot{a}\dot{b}}^{ij}\theta^{\dot{b}}\right)$.
All of relevant properties follow from the $SO\left(8\right)$ Fierz
identities derived from \eqref{eq:so8gammacommutations}, \emph{e.g.}
\begin{eqnarray}
\sigma_{a\dot{a}}^{i}\sigma_{\dot{b}\dot{c}}^{ij} & = & -\sigma_{a\dot{b}}^{i}\sigma_{\dot{a}\dot{c}}^{ij}+2\eta_{\dot{a}\dot{b}}\sigma_{a\dot{c}}^{j}-\eta_{\dot{b}\dot{c}}\sigma_{a\dot{a}}^{j}-\eta_{\dot{a}\dot{c}}\sigma_{a\dot{b}}^{j}.
\end{eqnarray}

In complete analogy, one is able to construct the massless light-cone
vertices with momentum $P^{-}=\sqrt{2}k$ and $P^{+}=0$. The result
is:\begin{subequations}\label{eq:masslessverticesp-}
\begin{eqnarray}
\overline{U}_{i}\left(z;k\right) & \equiv & e^{-ik\sqrt{2}X_{L}^{+}}\left\{ \Lambda_{i}+\left(\frac{k}{3!}\right)\theta_{ij}\Lambda_{j}+\left(\frac{k^{2}}{5!}\right)\theta_{ij}\theta_{jk}\Lambda_{k}+\left(\frac{k^{3}}{7!}\right)\theta_{ij}\theta_{jk}\theta_{kl}\Lambda_{l}\right\} ,\\
\overline{Y}_{\dot{a}}\left(z;k\right) & \equiv & e^{-ik\sqrt{2}X_{L}^{+}}\left(\theta\sigma^{i}\right)_{\dot{a}}\left\{ \left(\frac{1}{2!}\right)\Lambda_{i}+\left(\frac{k}{4!}\right)\theta_{ij}\Lambda_{j}+\left(\frac{k^{2}}{6!}\right)\theta_{ij}\theta_{jk}\Lambda_{k}+\left(\frac{k^{3}}{8!}\right)\theta_{ij}\theta_{jk}\theta_{kl}\Lambda_{l}\right\} \nonumber \\
 &  & +e^{-ik\sqrt{2}X_{L}^{+}}\left(\frac{\lambda_{\dot{a}}}{k}\right),
\end{eqnarray}
\end{subequations}where $\Lambda^{i}\equiv\left(\theta^{a}\sigma_{a\dot{a}}^{i}\lambda^{\dot{a}}\right)$
and $\theta^{ij}\equiv\left(\theta^{a}\sigma_{ab}^{ij}\theta^{b}\right)$.
Introducing the polarizations of the massless vector, $a^{i}$, and
spinor, $\xi^{\dot{a}}$, the unintegrated vertex operator in this
case can be cast as $\lambda^{\dot{a}}A_{\dot{a}}=a^{i}\overline{U}_{i}+\xi^{\dot{a}}\overline{Y}_{\dot{a}}$,
with
\begin{eqnarray}
A_{\dot{a}} & = & e^{-ik\sqrt{2}X_{L}^{+}}\left\{ \delta_{il}+\left(\frac{k}{3!}\right)\theta_{il}+\left(\frac{k^{2}}{5!}\right)\theta_{ij}\theta_{jl}+\left(\frac{k^{3}}{7!}\right)\theta_{ij}\theta_{jk}\theta_{kl}\right\} a^{i}\left(\sigma^{l}\theta\right)_{\dot{a}}+e^{-ik\sqrt{2}X_{L}^{+}}\left(\frac{\xi_{\dot{a}}}{k}\right)\nonumber \\
 &  & +e^{-ik\sqrt{2}X_{L}^{+}}\left\{ \left(\frac{1}{2!}\right)\delta_{il}+\left(\frac{k}{4!}\right)\theta_{il}+\left(\frac{k^{2}}{6!}\right)\theta_{ij}\theta_{jl}+\left(\frac{k^{3}}{8!}\right)\theta_{ij}\theta_{jk}\theta_{kl}\right\} \left(\xi\sigma^{i}\theta\right)\left(\sigma^{l}\theta\right)_{\dot{a}}.\label{eq:lightcone-superspinor}
\end{eqnarray}

The next step is to derive the expressions for the superfields of
\eqref{eq:superfields}. Since the construction is very similar for
$P^{+}\neq0$ and $P^{-}\neq0$, the latter will be used in order
to illustrate the procedure. It may be useful to emphasize that $A_{a}=0$
and $D_{\dot{a}}A_{\dot{b}}=0$ (the dependence on $\theta_{\dot{a}}$
was removed) due to a gauge choice. Besides, $D_{a}A_{\dot{a}}=iA_{i}\sigma_{a\dot{a}}^{i}$.
This can be seen from the Fierz decomposition of $D_{a}A_{\dot{a}}$,
given by
\begin{equation}
-i\left(D_{a}A_{\dot{a}}\right)=A_{i}\sigma_{a\dot{a}}^{i}+A_{ijk}\sigma_{a\dot{a}}^{ijk}.
\end{equation}
The last term is proportional to the linearized equation of motion
for $A_{\alpha}$, $\left(D\gamma^{+-ijk}A\right)=0$, so $A_{ijk}=0$.
The first term, $A^{i}$, represents the non-vanishing vector components
of the superfield given in \eqref{eq:supervector}:
\begin{eqnarray}
A_{i} & = & e^{-ik\sqrt{2}X_{L}^{+}}\left\{ \delta_{ij}+\left(\frac{k}{2!}\right)\theta_{ji}+\left(\frac{k^{2}}{4!}\right)\theta_{jk}\theta_{ki}+\left(\frac{k^{3}}{6!}\right)\theta_{jl}\theta_{lk}\theta_{ki}+\left(\frac{k^{4}}{8!}\right)\theta_{jm}\theta_{mk}\theta_{kl}\theta_{li}\right\} a^{j}\nonumber \\
 &  & +e^{-ik\sqrt{2}X_{L}^{+}}\left\{ \delta_{ij}+\left(\frac{k}{3!}\right)\theta_{ji}+\left(\frac{k^{2}}{5!}\right)\theta_{jk}\theta_{ki}+\left(\frac{k^{3}}{7!}\right)\theta_{jl}\theta_{lk}\theta_{ki}\right\} \left(\xi\sigma^{j}\theta\right).\label{eq:lightcone-supervector}
\end{eqnarray}

For $W^{\alpha}$, the $SO\left(8\right)$ decomposition of \eqref{eq:superspinor}
gives $W^{a}=0$ and $W^{\dot{a}}=-ikA^{\dot{a}}$. At last, the non-vanishing
components of the super field strength are $\sqrt{2}F_{+i}=-\sqrt{2}F_{i+}=-ikA_{i}$,
completing all the blocks needed for the construction of the integrated
massless vertex. Notice that the particular gauge of the approach
presented here enables the simple component expansion of the $SO\left(8\right)$
superfields of \eqref{eq:lightcone-superspinor} and \eqref{eq:lightcone-supervector},
which were already discussed in the work of Brink, Green and Schwarz
\cite{Brink:1983pf}.

\section*{DDF operators: definition and algebra}

The gauge fixed version of \eqref{eq:integratedmassless} is given
by
\begin{equation}
\overline{V}_{\textrm{L.C.}}\left(k;a^{i},\xi^{\dot{a}}\right)=\frac{1}{2\pi i}\oint\left\{ \left(\Pi_{i}-i\sqrt{2}k\overline{N}_{i}\right)A^{i}+\left(i\partial\theta^{\dot{a}}+kd^{\dot{a}}\right)A_{\dot{a}}\right\} ,\label{eq:LCintegratedvertex}
\end{equation}
where
\begin{equation}
\overline{N}^{i}\equiv N^{+i}=-\frac{1}{\sqrt{2}}\omega^{\dot{a}}\lambda^{a}\sigma_{a\dot{a}}^{i}.
\end{equation}
As a consistency check, observe that
\begin{equation}
\left[Q_{\textrm{BRST}},\overline{V}_{\textrm{L.C.}}\right]=-\frac{1}{2\pi i}\oint\left\{ \partial\left(\lambda^{\dot{a}}A_{\dot{a}}\right)+\sqrt{2}k^{2}\overline{N}^{i}\left(\lambda^{a}\sigma_{a\dot{a}}^{i}A^{\dot{a}}\right)\right\} .
\end{equation}
The first term inside the curly brackets is a total derivative whereas
the last one vanishes due to the pure spinor constraint $\lambda^{a}\lambda_{a}=0$,
cf. equation \eqref{eq:PS-SO8}, as
\begin{eqnarray*}
\overline{N}^{i}\left(\lambda\sigma^{i}\right)_{\dot{a}} & = & -\frac{1}{\sqrt{2}}\omega^{\dot{b}}\lambda^{a}\lambda^{b}\left(\sigma_{a\dot{a}}^{i}\sigma_{b\dot{b}}^{i}\right)\\
 & = & -\frac{1}{2\sqrt{2}}\omega^{\dot{b}}\lambda^{a}\lambda^{b}\left(\sigma_{a\dot{a}}^{i}\sigma_{b\dot{b}}^{i}+\sigma_{a\dot{b}}^{i}\sigma_{b\dot{b}}^{i}\right)\\
 & = & -\frac{1}{\sqrt{2}}\omega_{\dot{a}}\left(\lambda^{a}\lambda_{a}\right).
\end{eqnarray*}
Hence $\left[Q_{\textrm{BRST}},\overline{V}_{\textrm{L.C.}}\right]=0$.
It might be helpful to point out that $\partial=\Pi^{+}\partial_{+}-i\partial\theta^{a}D_{a}$
whenever acting on superfields that depend only on $X^{+}$ and $\theta^{a}$.

Defining the DDF operators $\overline{V}_{i}$ and $\overline{W}_{\dot{a}}$
through
\begin{equation}
\overline{V}_{\textrm{L.C.}}\left(k;a^{i},\xi^{\dot{a}}\right)\equiv a^{i}\overline{V}_{i}\left(k\right)-i\xi^{\dot{a}}\overline{W}_{\dot{a}}\left(k\right),\label{eq:defDDF}
\end{equation}
it will be demonstrated that\begin{subequations}\label{eq:DDFalgebra}
\begin{eqnarray}
\left[\overline{V}_{i}\left(k\right),\overline{V}_{j}\left(p\right)\right] & = & \sqrt{2}k\delta_{ij}\delta_{p+k}P^{+},\\
\left[\overline{V}_{i}\left(k\right),\overline{W}_{\dot{a}}\left(p\right)\right] & = & 0,\\
\left\{ \overline{W}_{\dot{a}}\left(k\right),\overline{W}_{\dot{b}}\left(p\right)\right\}  & = & \sqrt{2}\delta_{\dot{a}\dot{b}}\delta_{p+k}P^{+},
\end{eqnarray}
\end{subequations}which consists of a creation/annihilation algebra
whenever acting on states with $P^{+}\neq0$.

Although the following demonstration does not rely on the explicit
computation of the (anti)commutators in \eqref{eq:DDFalgebra}, the
explicit form of the DDF operators will be presented here for completeness:\begin{subequations}
\begin{eqnarray}
\overline{V}_{i}\left(k\right) & = & \frac{1}{2\pi i}\oint\left\{ \Pi_{i}-i\sqrt{2}k\overline{N}_{i}+\left(i\partial\theta^{\dot{a}}+kd^{\dot{a}}\right)\left(\sigma_{i}\theta\right)_{\dot{a}}\right\} \\
 &  & +\frac{1}{2\pi i}\oint\left\{ \left(\frac{k}{2!}\right)\theta_{ij}+\left(\frac{k^{2}}{4!}\right)\theta_{ik}\theta_{kj}+\left(\frac{k^{3}}{6!}\right)\theta_{il}\theta_{lk}\theta_{kj}+\left(\frac{k^{4}}{8!}\right)\theta_{im}\theta_{mk}\theta_{kl}\theta_{lj}\right\} \Pi_{j}e^{-ik\sqrt{2}X_{L}^{+}}\nonumber \\
 &  & -\frac{k}{\sqrt{2}\pi}\oint\left\{ \left(\frac{k}{2!}\right)\theta_{ij}+\left(\frac{k^{2}}{4!}\right)\theta_{ik}\theta_{kj}+\left(\frac{k^{3}}{6!}\right)\theta_{il}\theta_{lk}\theta_{kj}+\left(\frac{k^{4}}{8!}\right)\theta_{im}\theta_{mk}\theta_{kl}\theta_{lj}\right\} \overline{N}_{j}e^{-ik\sqrt{2}X_{L}^{+}}\nonumber \\
 &  & +\frac{1}{2\pi i}\oint\left\{ \left(\frac{k}{3!}\right)\theta_{il}+\left(\frac{k^{2}}{5!}\right)\theta_{ij}\theta_{jl}+\left(\frac{k^{3}}{7!}\right)\theta_{ij}\theta_{jk}\theta_{kl}\right\} \left(i\partial\theta^{\dot{a}}+kd^{\dot{a}}\right)\left(\sigma^{l}\theta\right)_{\dot{a}}e^{-ik\sqrt{2}X_{L}^{+}},\nonumber \\
\overline{W}_{\dot{a}}\left(k\right) & = & \frac{1}{2\pi}\oint\left\{ -d^{\dot{a}}+\sqrt{2}\partial X^{+}\theta^{\dot{a}}+\Pi_{i}\left(\sigma^{i}\theta\right)_{\dot{a}}+\frac{i}{2}\left(\sigma^{i}\theta\right)_{\dot{a}}\left(\partial\theta^{\dot{c}}\sigma_{c\dot{c}}^{i}\theta^{c}\right)\right\} e^{-ik\sqrt{2}X_{L}^{+}}\\
 &  & +\frac{1}{2\pi}\oint\left\{ \left(\frac{k}{3!}\right)\theta_{ji}+\left(\frac{k^{2}}{5!}\right)\theta_{jk}\theta_{ki}+\left(\frac{k^{3}}{7!}\right)\theta_{jl}\theta_{lk}\theta_{ki}\right\} \left(\sigma^{j}\theta\right)_{\dot{a}}\Pi_{i}e^{-ik\sqrt{2}X_{L}^{+}}\nonumber \\
 &  & +\frac{k\sqrt{2}}{2\pi i}\oint\left\{ \delta_{ij}+\left(\frac{k}{3!}\right)\theta_{ji}+\left(\frac{k^{2}}{5!}\right)\theta_{jk}\theta_{ki}+\left(\frac{k^{3}}{7!}\right)\theta_{jl}\theta_{lk}\theta_{ki}\right\} \left(\sigma^{j}\theta\right)_{\dot{a}}\overline{N}_{i}e^{-ik\sqrt{2}X_{L}^{+}}\nonumber \\
 &  & -\frac{1}{2\pi i}\oint\left\{ \left(\frac{k}{4!}\right)\theta_{il}+\left(\frac{k^{2}}{6!}\right)\theta_{ij}\theta_{jl}+\left(\frac{k^{3}}{8!}\right)\theta_{ij}\theta_{jk}\theta_{kl}\right\} \left(\sigma^{i}\theta\right)_{\dot{a}}\left(\partial\theta^{\dot{c}}\sigma_{c\dot{c}}^{l}\theta^{c}\right)e^{-ik\sqrt{2}X_{L}^{+}}\nonumber \\
 &  & +\frac{k}{2\pi}\oint\left\{ \left(\frac{1}{2!}\right)\delta_{il}+\left(\frac{k}{4!}\right)\theta_{il}+\left(\frac{k^{2}}{6!}\right)\theta_{ij}\theta_{jl}+\left(\frac{k^{3}}{8!}\right)\theta_{ij}\theta_{jk}\theta_{kl}\right\} \left(\sigma^{i}\theta\right)_{\dot{a}}\left(d^{\dot{c}}\sigma_{c\dot{c}}^{l}\theta^{c}\right)e^{-ik\sqrt{2}X_{L}^{+}}.\nonumber
\end{eqnarray}
\end{subequations} Observe that $\overline{V}_{j}\left(k\right)^{\dagger}=\overline{V}_{j}\left(-k\right)$
and $\overline{W}_{\dot{a}}\left(k\right)^{\dagger}=\overline{W}_{\dot{a}}\left(-k\right)$,
as expected. Besides $\overline{V}_{j}\left(0\right)=-iP_{j}$ and
$\overline{W}_{\dot{a}}\left(0\right)=q_{\dot{a}}$.

Given two vertices $V_{\textrm{L.C.}}$ with polarizations $\left(a^{i},\xi^{\dot{a}}\right)$
and $\left(b^{i},\chi^{\dot{a}}\right)$, their commutator is computed
to be
\begin{eqnarray}
\left[\overline{V}_{\textrm{L.C.}}\left(k;a^{i},\xi^{\dot{a}}\right),\overline{V}_{\textrm{L.C.}}\left(p;b^{i},\chi^{\dot{a}}\right)\right] & = & -\frac{k}{2\pi i}\oint\left\{ A_{\dot{a}}\left(k,a,\xi\right)\partial A_{\dot{a}}\left(p,b,\chi\right)\right\} \nonumber \\
 &  & +\frac{1}{2\pi i}\oint\left\{ A_{i}\left(k,a,\xi\right)\partial A^{i}\left(p,b,\chi\right)\right\} \nonumber \\
 &  & -\frac{ik}{2\pi i}\oint\left\{ A_{\dot{a}}\left(k,a,\xi\right)\partial\theta^{a}D_{a}A_{\dot{a}}\left(p,b,\chi\right)\right\} ,\label{eq:verticescomm}
\end{eqnarray}
and vanishes for $\left(k+p\right)\neq0$ , as the integrand is a
total derivative:
\begin{multline}
A_{i}\left(k\right)\partial A_{i}\left(p\right)-kA_{\dot{a}}\left(k\right)\partial A_{\dot{a}}\left(p\right)-ikA_{\dot{a}}\left(k\right)\partial\theta^{a}\left[D_{a}A_{\dot{a}}\left(p\right)\right]=\\
=\frac{p}{k+p}\partial\left\{ A_{i}\left(k\right)A_{i}\left(p\right)-kA_{\dot{a}}\left(k\right)A_{\dot{a}}\left(p\right)\right\} .\label{eq:totder}
\end{multline}

Setting $\left(k+p\right)$ to zero (with $k\neq0$), the expression
inside the curly brackets in \eqref{eq:totder} is a constant written
in terms of the polarizations, directly shown to be
\begin{equation}
A_{i}\left(k,a,\xi\right)A_{i}\left(-k,b,\chi\right)-kA_{\dot{a}}\left(k,a,\xi\right)A_{\dot{a}}\left(-k,b,\chi\right)=a_{i}b_{i}+\left(\frac{1}{k}\right)\xi_{\dot{a}}\chi_{\dot{a}}.\label{eq:SFconstraint}
\end{equation}
Note also that
\begin{eqnarray}
A_{i}\left(k\right)\partial\theta^{a}D_{a}A_{i}\left(-k\right) & = & -ik\left(\theta^{a}\partial\theta_{a}\right)\left(A_{i}\left(k\right)A_{i}\left(-k\right)-kA_{\dot{a}}\left(k\right)A_{\dot{a}}\left(-k\right)\right)\nonumber \\
 &  & -i\frac{k}{2}\left(\partial\theta_{ij}\right)\left(A^{i}\left(k\right)A^{j}\left(-k\right)+\frac{k}{4}A_{\dot{a}}\left(k\right)A_{\dot{b}}\left(-k\right)\sigma_{\dot{a}\dot{b}}^{ij}\right).
\end{eqnarray}
The second line of this equation is a total derivative and using this
result together with \eqref{eq:totder} and \eqref{eq:SFconstraint},
the right-hand side of \eqref{eq:verticescomm} is rewritten in a
very simple manner,
\begin{equation}
\left[\overline{V}_{\textrm{L.C.}}\left(k;a^{i},\xi^{\dot{a}}\right),\overline{V}_{\textrm{L.C.}}\left(p;b^{i},\chi^{\dot{a}}\right)\right]=\delta_{p+k}\sqrt{2}\left(ka_{i}b_{i}+\xi_{\dot{a}}\chi_{\dot{a}}\right)P^{+},\label{eq:fullDDFcommutator}
\end{equation}
which implies the DDF algebra of \eqref{eq:DDFalgebra} according
to the definition \eqref{eq:defDDF}.

In \cite{Mukhopadhyay:2005zu}, using a different gauge for the massless
vertices, the DDF-algebra was determined up to $\mathcal{O}\left(\theta^{2}\right)$.
The DDF operators are BRST-closed by construction, so are their (anti)commutators.
Since there is no other candidate to compose the massless pure spinor
cohomology, $\mathcal{O}\left(\theta^{2}\right)$ must be BRST-exact
and decouples naturally in the definition of the orthonormal basis
presented by Mukhopadhyay. Another way to present this argument is
recording that the difference between the DDF-operators of \eqref{eq:defDDF}
and the ones presented in \cite{Mukhopadhyay:2005zu} is a simple
BRST transformation (a gauge choice), implying that $\mathcal{O}\left(\theta^{2}\right)$
is BRST-trivial.

\section*{Physical spectrum}

The closed string case will be discussed first. To understand the
action of the operators $\overline{V}_{i}$ and $\overline{W}_{\dot{a}}$,
consider the commutator $\left[\overline{V}_{i}\left(p\right),U_{j}\left(z;k\right)\right]$.
The OPE
\begin{equation}
e^{-ip\sqrt{2}X_{L}^{+}}\left(y\right)e^{-ik\sqrt{2}X_{L}^{-}}\left(z\right)\sim\left(y-z\right)^{-2\left(k\cdot p\right)}:e^{-i\sqrt{2}\left(pX_{L}^{+}+kX_{L}^{-}\right)}:+\ldots\label{eq:OPEclosedX+X-}
\end{equation}
will always appear and it is directly related to the mass levels of
the superstring. Single-valuedness of \eqref{eq:OPEclosedX+X-} will
imply the discretization of $p$ in terms of $k$: $2\left(k\cdot p\right)\in\mathbb{Z}$.
This will be required in order for the operators to have any meaning
at all%
\footnote{This is of course frame independent, but the light-cone frame used
here is much easier to deal with, as unphysical polarizations (BRST-exact
states) are straightforward to identify. The physical meaning of this
quantization is deeply related to the Virasoro conditions, as it is
clear in the bosonic string, for example. However, the gauge-fixing
mechanism that provides the pure spinor formalism its BRST-like charge
was just recently discovered \cite{Berkovits:2014aia}, where a twistor
like constraint replaces the usual Virasoro ones. That is why this
quantization condition seems to be technical rather than fundamentally
based. %
}, determining an acceptable operation for $\overline{V}_{i}$ and
$\overline{W}_{\dot{a}}$ when commuting with $U_{j}$ and $Y_{a}$.
Note also that whenever $k\cdot p\leq0$, the commutator vanishes,
as there are only simple poles coming from the contractions of $d_{\dot{a}}$
with the superfields $U_{j}$ and $Y_{a}$.

From the state-operator map, the ground states associated to the algebra
\eqref{eq:DDFalgebra} will be denoted by $\left|i;k\right\rangle $
and $\left|a;k\right\rangle $. They correspond, respectively, to
the operators $U{}_{i}\left(z;k\right)$ and $Y_{a}\left(z;k\right)$
defined in \eqref{eq:masslessverticesp+}. Then, the DDF-operators
$\overline{V}_{i}\left(\frac{n}{2k}\right)$ and $\overline{W}_{\dot{a}}\left(\frac{n}{2k}\right)$,
where $n\in\mathbb{Z}^{+}$, will generate the excited configurations.
For example,
\[
\begin{array}{cc}
\overline{V}_{i}\left(\frac{1}{2k}\right)\left|j;k\right\rangle , & \overline{W}_{\dot{a}}\left(\frac{1}{2k}\right)\left|a;k\right\rangle ,\\
\overline{V}_{i}\left(\frac{1}{2k}\right)\left|a;k\right\rangle , & \overline{W}_{\dot{a}}\left(\frac{1}{2k}\right)\left|j;k\right\rangle ,
\end{array}
\]
are the first massive level of the holomorphic sector of the closed
string. Clearly, the physical states are the direct (level-matched)
product of the holomorphic and anti-holomorphic sectors, and the full
vertex operators generated through this procedure will have momentum
$P^{+}=-\sqrt{2}k$ and $P^{-}=-\frac{N}{\sqrt{2}k}$ (with $N\in\mathbb{Z}^{*}$),
satisfying the mass-shell condition
\begin{equation}
m_{\textrm{closed}}^{2}=2P^{+}P^{-}=2N.\label{eq:closedmassshell}
\end{equation}

Concerning supersymmetry, it is easy to demonstrate that
\begin{equation}
\begin{array}{rclrcl}
\left\{ q_{a},U_{i}\left(k\right)\right\}  & = & 0, & \left[q_{a},\overline{V}_{i}\left(k\right)\right] & = & -ik\sigma_{a\dot{a}}^{i}\overline{W}_{\dot{a}}\left(k\right),\\
\left\{ q_{\dot{a}},U^{i}\left(k\right)\right\}  & = & k\sigma_{a\dot{a}}^{i}Y^{a}\left(k\right), & \left[q_{\dot{a}},\overline{V}_{i}\left(k\right)\right] & = & 0,\\
\left[q_{a},Y_{b}\left(k\right)\right] & = & 0, & \left\{ q_{a},\overline{W}_{\dot{a}}\left(k\right)\right\}  & = & i\sigma_{a\dot{a}}^{i}\overline{V}_{i}\left(k\right),\\
\left[q_{\dot{a}},Y_{a}\left(k\right)\right] & = & \sigma_{a\dot{a}}^{i}U_{i}\left(k\right), & \left\{ q_{\dot{a}},\overline{W}_{\dot{b}}\left(k\right)\right\}  & = & \sqrt{2}\eta_{\dot{a}\dot{b}}\delta_{k}P^{+},
\end{array}\label{eq:susycreationalgebra}
\end{equation}
and the supersymmetric structure of the spectrum is trivially shown.

For the open string, the steps are almost the same. The difference
arises in the definition of the integrated vertex operators, which
will now be given in terms of an integral on the boundary of the disk.
The analogue of the OPE \eqref{eq:OPEclosedX+X-} is
\begin{equation}
e^{ip\sqrt{2}X_{L}^{+}}\left(z\right)e^{ik\sqrt{2}X_{L}^{-}}\left(y\right)\sim\left(z-y\right)^{-4\left(k\cdot p\right)}{}_{*}^{*}e^{i\sqrt{2}\left(pX_{L}^{+}+kX_{L}^{-}\right)}{}_{*}^{*}+\ldots,\label{eq:OPEopenX+X-}
\end{equation}
where the operators are now \emph{boundary} \emph{normal ordered }$_{*}^{*}\:_{*}^{*}$.
The single-valuedness condition will now be $8\left(k\cdot p\right)\in\mathbb{Z}$,
as only half of the complex plane appears in the definition of the
line integral. In other words, while the worldsheet coordinate $\sigma\in\left[0,2\pi\right)$
for the closed string, in the open string $\sigma\in\left[0,\pi\right]$.
Then, the mass levels of the open string will be
\begin{equation}
m_{\textrm{open}}^{2}=2P^{+}P^{-}=\frac{N}{2}.\label{eq:openmassshell}
\end{equation}

Therefore, the light-cone unintegrated vertex operators in the pure
spinor superstring at any mass level can all be manufactured from
chains of commutators between the DDF operators with $p=\frac{n}{2k}$
($n>0$) and the massless (ground) operators $U{}_{i}\left(z;k\right)$
or $Y_{a}\left(z;k\right)$. Being BRST-closed by construction, this
shows that the cohomology includes the light-cone superstring spectrum.
An interesting observation is that the generated states contain only
half of the $\lambda^{\alpha}$ components, namely $\lambda^{a}$,
as it is the one that appears in $U{}_{i}$ and $Y_{a}$, and the
pure spinor contribution to $\overline{V}_{i}\left(p\right)$ and
$\overline{W}_{\dot{a}}\left(p\right)$ comes from the Lorentz current
$\overline{N}^{i}=-\frac{1}{\sqrt{2}}\lambda^{a}\omega^{\dot{a}}\sigma_{a\dot{a}}^{i}$.
Clearly, if the ground states were chosen to be $\overline{U}{}_{i}\left(z;k\right)$
and $\overline{Y}_{\dot{a}}\left(z;k\right)$ in \eqref{eq:masslessverticesp-},
the DDF operators would have been built out of the massless integrated
vertices with $P^{+}\neq0$ and the light-cone spectrum would only
depend on $\lambda^{\dot{a}}$.

The versatility of the spectrum generating algebra also extends to
the integrated vertex operators. In order to see that, the $P^{+}\neq0$
version of the DDF-operators will be needed and denoted by $V_{i}\left(k\right)$
and $W_{a}\left(k\right)$. The procedure is completely analogous
to the one presented above, just replacing $U{}_{i}\left(z;k\right)$
and $Y_{a}\left(z;k\right)$ by $V_{i}\left(k\right)$ and $W_{a}\left(k\right)$,
respectively. Note that their explicit form can be easily recovered
from \eqref{eq:LCintegratedvertex}, by defining\begin{subequations}
\begin{eqnarray}
V_{\textrm{L.C.}}\left(k;a^{i},\xi^{a}\right) & = & \frac{1}{2\pi i}\oint\left\{ \left(\Pi_{i}-i\sqrt{2}kN_{i}\right)A^{i}+\left(i\partial\theta^{a}+kd^{a}\right)A_{a}\right\} \\
 & \equiv & a^{i}V_{i}\left(k\right)-i\xi^{a}W_{a}\left(k\right),
\end{eqnarray}
\end{subequations}where $A_{i}$ and $A_{a}$ are the $SO\left(8\right)$-covariant
superfields in the frame $P^{-}=P^{i}=0$ and $P^{+}=\sqrt{2}k$ constructed
out of \eqref{eq:masslessverticesp+}. There is now a subtlety that
comes from the operators $\overline{V}_{i}\left(\frac{-1}{2k}\right)$
and $\overline{W}_{\dot{a}}\left(\frac{-1}{2k}\right)$, since the
simple pole argument used for the annihilation of the unintegrated
vertices when $k\cdot p<0$ does not work anymore. In other words,
the commutators
\begin{equation}
\begin{array}{cc}
\left[\overline{V}_{i}\left(\frac{-1}{2k}\right),V_{j}\left(k\right)\right], & \left[\overline{W}_{\dot{a}}\left(\frac{-1}{2k}\right),V_{j}\left(k\right)\right],\\
\left[\overline{V}_{i}\left(\frac{-1}{2k}\right),W_{a}\left(k\right)\right], & \left\{ \overline{W}_{\dot{a}}\left(\frac{-1}{2k}\right),W_{a}\left(k\right)\right\} ,
\end{array}\label{eq:inttachyon}
\end{equation}
are no longer guaranteed to vanish, which is potentially dangerous
as they would imply physical vertices corresponding to states with
$m^{2}<0$, that is, tachyonic states.

In the pure spinor formalism, the vanishing of the vertices \eqref{eq:inttachyon}
is related to the level of the ghost Lorentz algebra. Basically, the
relevant OPE that will appear in the computation is
\begin{equation}
\overline{N}^{i}\left(z\right)N^{j}\left(y\right)\sim-3\frac{\eta^{ij}}{\left(z-y\right)^{2}}-\frac{N^{ij}-\eta^{ij}N}{\left(z-y\right)},
\end{equation}
and precisely the factor of $-3$ ensures the absence of tachyons
in this DDF description. In other words, the level of the ghost Lorentz
algebra implies that there is no simple pole in the OPE between the
integrands of $V_{\textrm{L.C.}}$ and $\overline{V}_{\textrm{L.C.}}$
when \eqref{eq:inttachyon} is concerned. It is interesting to note
that the pure spinor ghosts play no role at all in the derivation
of the DDF algebra \eqref{eq:DDFalgebra}, which means that the light-cone
version of the massless vertex proposed by Siegel obeys the same algebra
and the vertices suggested in \eqref{eq:inttachyon} would exist.
Roughly speaking, the ``incompleteness'' of the vertex
\begin{equation}
V_{\textrm{Siegel}}=\frac{1}{2\pi i}\oint\left\{ \Pi^{m}A_{m}+i\partial\theta^{\alpha}A_{\alpha}+id_{\alpha}W^{\alpha}\right\} ,
\end{equation}
leads to the existence of a tachyon in the physical spectrum. Perhaps
a clearer example is the DDF construction in the bosonic string. Defining\begin{subequations}
\begin{eqnarray}
V_{\textrm{bos}}^{i}\left(k\right) & \equiv & \frac{1}{2\pi i}\oint\partial X_{i}e^{-ik\sqrt{2}X_{L}^{-}},\\
\overline{V}_{\textrm{bos}}^{i}\left(k\right) & \equiv & \frac{1}{2\pi i}\oint\partial X_{i}e^{-ik\sqrt{2}X_{L}^{+}},
\end{eqnarray}
\end{subequations}it is easy to see the emergence of the tachyon
vertex operator:
\begin{equation}
\left[\overline{V}_{\textrm{bos}}^{i}\left(\frac{-1}{2k}\right),V_{\textrm{bos}}^{j}\left(k\right)\right]=-\frac{\eta^{ij}}{2\pi i}\underbrace{\oint\exp\left\{ -ik\sqrt{2}X_{L}^{-}+\frac{i}{\sqrt{2}k}X_{L}^{+}\right\} }_{\propto\: V_{\textrm{tachyon}}}.
\end{equation}

The level of the Lorentz current can be argued to be an evidence of
the quantum equivalence of the pure spinor vertex operator \eqref{eq:integratedmassless}
with the RNS massless one \cite{Berkovits:2000fe}. The spectrum generating
algebra presented here, more than supporting this equivalence, shows
through a simple construction that the light-cone spectrum of the
pure spinor superstring coincides with the ones of the RNS and the
Green-Schwarz formalisms. It must be pointed out that this is not
a demonstration that the DDF-states span the pure spinor cohomology.
For further details, see \cite{Berkovits:2014bra}.

\section*{Perspectives}

In a deeper analogy with the RNS string \cite{Brower:1973}, it might
be interesting to investigate whether a full DDF algebra will also
exist here. If this is the case, it is expected that Siegel's like
constraints will appear as symmetry generators for the operators $\overline{V}_{i}$
and $\overline{W}_{\dot{a}}$ (resembling, for example, the action
of $\overline{V}_{\textrm{bos}}^{-}$ in the usual bosonic construction,
that satisfy $\left[\overline{V}_{\textrm{bos}}^{-}\left(p\right),\overline{V}_{\textrm{bos}}^{i}\left(k\right)\right]\propto\overline{V}_{\textrm{bos}}^{i}\left(p+k\right)$).
A better understanding of the cohomology structure of the pure spinor
formalism may help clarify its equivalence with the RNS and Green-Schwarz
superstrings, which is now far from evident.

The flat space construction of the spectrum generating algebra is
also very appealing for it allows the determination of the whole physical
spectrum by knowing just the massless vertex operators. It would be
great if this same procedure could shed some light in the current
knowledge about the superstring spectrum in curved backgrounds. Recently,
the massless cohomology of the pure spinor superstring in $AdS_{5}\times S^{5}$
was determined in the limit close to the $AdS$ boundary \cite{Berkovits:2012},
providing some ground for further studies on the generalization of
the results presented here.

\section*{Acknowledgements}

This work was motivated by a parallel research project on a light-cone
analysis of the pure spinor formalism \cite{Berkovits:2014bra}. I
would like to thank Nathan Berkovits for very useful suggestions.
I would like to thank also FAPESP grants 2009/17516-4 for financial
support and 2011/11973-4 (ICTP-SAIFR) for providing a stimulating
research environment.


\begin{thebibliography}{10}
\bibitem{Berkovits:2000fe}N.~Berkovits, ``Super Poincare covariant quantization of the superstring'', JHEP {\bf 0004}, 018 (2000) [hep-th/0001035].

\bibitem{Berkovits:2005ng}N.~Berkovits and C.~R.~Mafra, ``Equivalence of two-loop superstring amplitudes in the pure spinor and RNS formalisms,''   Phys.\ Rev.\ Lett.\  {\bf 96}, 011602 (2006)   [hep-th/0509234].

\bibitem{Berkovits:2002qx}N.~Berkovits and O.~Chandia, ``Massive superstring vertex operator in D = 10 superspace,''   JHEP {\bf 0208}, 040 (2002)   [hep-th/0204121].

\bibitem{Berkovits:2000nn}N.~Berkovits, ``Cohomology in the pure spinor formalism for the superstring'',   JHEP {\bf 0009}, 046 (2000)   [hep-th/0006003].

\bibitem{Berkovits:PSRNSGS}N.~Berkovits, ``Relating the RNS and pure spinor formalisms for the superstring'',   JHEP {\bf 0108}, 026 (2001)   [hep-th/0104247].

N.~Berkovits and D.~Z.~Marchioro, ``Relating the Green-Schwarz and pure spinor formalisms for the superstring'',   JHEP {\bf 0501}, 018 (2005)   [hep-th/0412198].

\bibitem{Del Giudice:1971fp}E.~Del Giudice, P.~Di Vecchia and S.~Fubini, ``General properties of the dual resonance model'',   Annals Phys.\  {\bf 70}, 378 (1972).

\bibitem{Mukhopadhyay:2005zu}P.~Mukhopadhyay, ``DDF construction and D-brane boundary states in pure spinor formalism,''   JHEP {\bf 0605}, 055 (2006)   [hep-th/0512161].

\bibitem{Siegel:1985xj}W.~Siegel, ``Classical Superstring Mechanics'', Nucl.\ Phys.\ B {\bf 263}, 93 (1986).

\bibitem{Berkovits:2014bra}N.~Berkovits and R.~L.~Jusinskas, ``Light-Cone Analysis of the Pure Spinor Formalism for the Superstring,''   arXiv:1406.2290 [hep-th].

\bibitem{Brink:1983pf}L.~Brink, M.~B.~Green and J.~H.~Schwarz, ``Ten-dimensional Supersymmetric {Yang-Mills} Theory With SO(8) - Covariant Light Cone Superfields'',   Nucl.\ Phys.\ B {\bf 223}, 125 (1983).

\bibitem{Brower:1973}R.~C.~Brower and K.~A.~Friedman, ``Spectrum Generating Algebra and No Ghost Theorem for the Neveu-schwarz Model'',   Phys.\ Rev.\ D {\bf 7}, 535 (1973).

\bibitem{Berkovits:2012}N.~Berkovits and T.~Fleury, ``Harmonic Superspace from the $AdS_5\times S^5$ Pure Spinor Formalism'',   JHEP {\bf 1303}, 022 (2013)   [arXiv:1212.3296].

\bibitem{Berkovits:2014aia}N.~Berkovits,  ``Twistor Origin of the Superstring,''   arXiv:1409.2510 [hep-th].\end{thebibliography}
\end{document}